\documentclass[epj]{svjour}
\usepackage{graphics}
\usepackage{float}
\usepackage{amsmath}
\usepackage{amssymb}
\usepackage{cancel}
\usepackage{stmaryrd}
\usepackage{graphicx}

\begin{document}
\title{Equivalence principle violation at finite temperature in scalar-tensor gravity}
\author{Massimo Blasone\inst{1,2,}\thanks{blasone@sa.infn.it} \and Salvatore Capozziello\inst{3,4,5}\thanks{capozziello@na.infn.it} \and Gaetano Lambiase\inst{1,2,}\thanks{lambiase@sa.infn.it} \and Luciano Petruzziello\inst{1,2,}\thanks{lpetruzziello@na.infn.it}
%
}                     
%
%
\institute{Dipartimento di Fisica, Universit\`a di Salerno, Via Giovanni Paolo II 132, I-84084, Fisciano (SA), Italy. \and INFN, Sezione di Napoli, Gruppo collegato di Salerno, Via Giovanni Paolo II 132, I-84084 Fisciano (SA), Italy. \and Dipartimento di Fisica, Universit\`a di Napoli ``Federico II'',
via Cinthia 9, I-80126, Napoli (Italy). \and Tomsk State Pedagogical University, ul. Kievskaya, 60, 634061 Tomsk (Russia).\and INFN, Sezione di Napoli,
via Cinthia 9, I-80126 Napoli (Italy).
}
\date{Received: date / Revised version: date}
%
\abstract{
We analyze possible violations of the Equivalence Principle  in scalar-tensor gravity  at finite temperature $T.$  Before we present an approach where the Equivalence Principle  violation is achieved within the framework of Quantum Field Theory. After,  we rely on an alternative approach first proposed by Gasperini, which leads to the same outcome obtained in the framework of Quantum Field Theory (one-loop corrections) at finite $T$. Finally, we exhibit the application of the above formalism both to a generic diagonal metric, cast in spherical coordinates, and to the Brans-Dicke theory. In the last case, we show that it is possible to put a significant constraint on the free parameter of the theory by means of experimental bounds on the Equivalence Principle.
} 
\maketitle
\section{Introduction}

The Equivalence Principle (EP) is one of the cornerstones of Einstein's General Relativity (GR). An example of its meaning can be found in the choice of the Riemannian geometry among all the available mathematical frameworks for the description of the bending of space-time due to gravity~\cite{mtw}. Furthermore, the validity of Equivalence Principle is based on the coincidence of the metric structure and the geodesic structure of the space-time.
Its relevance appears also  going beyond GR  in order to discriminate among competing  theories of gravity which one can be regarded as reliable. Moreover, the fact that it can be violated at  quantum regime is the evidence that there are still hidden issues to be discovered for a better understanding of EP in the context of quantum mechanics~\cite{zy}.

The first formulation of EP, also known as the weak equivalence
principle (WEP), states that the inertial mass $m_i$ and the gravitational mass
$m_g$ of any physical object are exactly the same. In other words, 
it is impossible to locally distinguish the effects due to a gravitational
field from the non-inertial consequences of a uniformly accelerated motion by means of a free-falling experiment~\cite{mtw}.
This awareness entails that it is always possible to locally describe a neighborhood of any folded space-time point with the language of Special Relativity. This is a crucial aspect; indeed, it is not surprising that Einstein himself used to address EP as \emph{``the happiest thought of my life''}.

More generally, WEP can be incorporated in a set of assumptions which goes by the name of Einstein equivalence principle (EEP), which states that:
\begin{itemize} 
\item [$\cdot$] WEP holds;
\item [$\cdot$] the result of any local non-gravitational test does not depend 
on the velocity of the free-falling experimental apparatus and on the position and the time in which it is carried out in the Universe.
\end{itemize}
By ``local non-gravitational test'' we mean an experiment that takes place 
in a small region of a free-falling laboratory. However, this implies neglecting the  gravitational interactions in the definition of WEP and EEP.
In order to account for modified theories of gravity, there is the need to introduce an even more general concept, which includes both the previous principles in a suitable limit. Such a requirement results in the strong equivalence principle (SEP), whose assumptions claim:
\begin{itemize}
\item [$\cdot$] WEP is valid for self-gravitating bodies as well as for test bodies;
\item [$\cdot$] the result of any local test does not depend on the velocity
of the free-falling experimental apparatus and on the position and the time in which it is carried out in the Universe.
\end{itemize}
Clearly, SEP reduces to EEP in the limit of vanishing gravitational field. 

In this work, our considerations are devoted to the  WEP, not only in the context of GR, but also for scalar-tensor gravity which is a particular case of Extended Theories of Gravity  (for a review, see Ref.~\cite{cap}, for applications Ref.~\cite{app}). For all the key points of the upcoming analysis, the role of temperature $T$ turns out to be crucial. Indeed, at zero temperature, EP still holds, because the contributions to $m_i$ and $m_g$ that allow $m_g/m_i\neq 1$ vanish as $T\rightarrow 0$. This result can be achieved by invoking two different approaches, which will be compared in the next sections. Thanks to one of them, the study can be generalized for any given metric with little effort, whereas the other one requires the evaluation of radiative corrections by Quantum Field Theory (QFT) techniques. 

The paper is organized as follows: in Sec.~II we briefly review some aspects of  scalar-tensor  gravity as particular cases of Extended Theories of Gravity. Sec.~III is devoted to the investigation of EP violation in the context of QFT at finite temperature, whereas, in Sec.~IV, the same outcome is produced by virtue of a different analysis, where the attention is focused on the modification of the geodesic equation in the case of Schwarzschild solution. With this formalism, we show that it is possible to extend the discussion to other physical environments, in which space-time can be described also by generic Extended Theories of Gravity. Sec.~VI is devoted to  conclusion and discussions.

Throughout the whole paper, we use the natural units $c=\hbar=1$ and the mostly-negative metric signature $\mathrm{diag}(+,-,-,-)$.

\section{Extended Theories of Gravity}
 Extended Theories of Ggravity are extensions of GR where  geometry  couples  to some scalar field and/or to higher-order curvature invariants. Such corrections  can appear in the Hilbert-Einstein action.  Combinations of scalar fields and higher order invariants  emerge in effective Lagrangians, producing mixed higher order/scalar-tensor gravity. As discussed in details in \cite{cap}, the paradigm is recovering GR at a certain regime so that we are dealing with {\it extensions} of GR and not with {\it alternatives} of GR\footnote{For example, teleparallel equivalent theories of gravity (TEGR) or other gauge theories are really alternative to GR. In the first case, dynamics is given by torsion instead of curvature and the space-time structure is related to tetrads instead of  metric \cite{cai}.}. 
Scalar-tensor gravity and $f(R)$ gravity  are the straightforward extensions of GR. Differently from scalar-tensor models, $f(R)$ gravity does not require
the introduction of further fields for the description of gravity but it is a straightforward generalization of the Hilbert-Einstein action where non-linear terms in the Ricci scalar $R$ are considered. It is worth noticing that $f(R)$ gravity can be considered under the standard of  scalar-tensor theories~\cite{cap,chi}. In fact, taking into account  conformal transformations, it is easy to see  that results holding for scalar-tensor gravity can be easily recast for $f(R)$ gravity. With this consideration in mind, in what follows, we will consider only the Brans-Dicke case.

\subsection{Scalar-tensor gravity}
Let us consider the  action~\cite{jor}
\begin{equation}
S_{J}=\int d^4x\sqrt{-g}\left[\varphi_{J}^{\gamma}\left(R-\omega_{J}\frac{1}{\varphi_{J}^{2}}g^{\mu\nu}\partial_{\mu}\varphi_{J}\partial_{\nu}\varphi_{J}\right)+\mathfrak{L}_\mathrm{matter}\left(\varphi_{J},\psi\right)\right],\label{eq:49}
\end{equation}
where $\varphi_{J}$ is the scalar field, $\gamma$ and $\omega_{J}$
are constants and $\psi$ contains the contribution of matter
fields. We immediately note that: 
\begin{itemize}
\item [$\cdot$] $-\omega_{J}\,{\varphi_{J}^{-2}}\,g^{\mu\nu}\partial_{\mu}\varphi_{J}\partial_{\nu}\varphi_{J}$
can be correctly interpreted as the kinetic contribution related
to the scalar field;
\vspace{1mm}
\item [$\cdot$] $\varphi_{J}^{\gamma}R$ is a non-minimal coupling term;
\vspace{1mm}
\item [$\cdot$] the Lagrangian density $\mathfrak{L}_{\mathrm{matter}}$ depends not only on the matter fields, but in principle, also on the scalar field.
\end{itemize}
The above expression describes a conspicuous number of models, according to the free parameters of the model. However, we are mainly concerned with the one developed by Brans and Dicke~\cite{bd}, which is examined in detail as follows.

\subsection{Brans-Dicke gravity}
Let us introduce the Brans-Dicke action~\cite{bd}, which is similar to the one showed in Eq.~(\ref{eq:49}),
but with several differences:
\begin{equation}
S_{BD}=\int d^4x\sqrt{-g}\left(\varphi R-\omega\frac{1}{\varphi}g^{\mu\nu}\partial_{\mu}\varphi\partial_{\nu}\varphi+\mathfrak{L}_\mathrm{matter}\left(\psi\right)\right).\label{eq:50}
\end{equation}
In Eq.~(\ref{eq:50}), the matter Lagrangian density does not depend
on the scalar field and $\gamma=1$. This is crucial,
because it means that $\omega$ is the only parameter of the theory.
Moreover, it is clear that:
\begin{equation}
\varphi=\frac{1}{16\pi G_{eff}},\label{eq:51}
\end{equation}
and such a result is traduced in the introduction of a new ``effective''
gravitational constant that has to be identified with the scalar field.
This consideration requires some restrictions.  In particular, it is essential
that $\varphi$ is spatially uniform, and it must vary slowly
with cosmic time. If these characteristics are not possessed by $\varphi$,
the theory cannot be consistent with experimental data, since they clearly support the presence of a gravitational constant that
enters field equations as GR predicts. On the other hand, experiments
may put a constraint on the only free parameter of the theory, namely $\omega$. In this sense, the Brans-Dicke gravity is genuinely an Extended Theory of Gravity which extends results of GR to be more "Machian" \cite{cap}.

Field equations derived from Eq.~(\ref{eq:50}) are given by:
\begin{equation}
2\varphi G_{\mu\nu}=T_{\mu\nu}+T_{\mu\nu}^{\varphi}-2\left(g_{\mu\nu}\boxempty-\nabla_{\mu}\nabla_{\nu}\right)\varphi,\label{eq:54}
\end{equation}
that can be obtained by means of a variation with respect to $g^{\mu\nu}$, and:
\begin{equation}
\boxempty\varphi=\zeta^{2}T,\label{eq:55}
\end{equation}
deduced by a variation with respect to $\varphi$, where $\zeta^{-2}=6+4\,\omega$
and $T=g^{\mu\nu}T_{\mu\nu}$. The symbol $\boxempty$ clearly denotes
the D'Alembert operator.
In Eq.~(\ref{eq:54}), $T_{\mu\nu}$ and $T_{\mu\nu}^{\varphi}$ are extracted
by varying $\mathfrak{L}_{matter}$ and the kinetic term of $S_{BD}$,
respectively.
As expected, field equations for the metric tensor becomes the ones derived by GR in the limit $\varphi=\mathrm{const}={1}/{16\pi G}$.

If a static and isotropic solution is now sought, it is possible
to find an expression for the line element~\cite{fuj}:
\begin{equation}
ds^{2}=e^{v}dt^{2}-e^{u}\left[dr^{2}+r^{2}\left(d\vartheta^{2}+sin^{2}\vartheta d\Phi^{2}\right)\right],\label{eq:eff57}
\end{equation}
where:
\begin{equation}
e^{v}=e^{2\alpha_{0}}\left(\frac{1-\frac{B}{r}}{1+\frac{B}{r}}\right)^{\frac{2}{\lambda}}, \qquad e^{u}=e^{2\beta_{0}}\left(1+\frac{B}{r}\right)^{4}\left(\frac{1-\frac{B}{r}}{1+\frac{B}{r}}\right)^{\frac{2\left(\lambda-C-1\right)}{\lambda}},\label{eq:eff58}
\end{equation}
with $\alpha_{0}$, $\beta_{0}$, $B$, $C$ and $\lambda$ being constants
that can be connected to the free parameter of the theory $\omega$.
Since it is a scalar-tensor theory, a solution for $\varphi$ must also
be found; in the considered case, the outcome turns out to be:
\begin{equation}
\varphi=\varphi_{0}\left(\frac{1-\frac{B}{r}}{1+\frac{B}{r}}\right)^{-\frac{C}{\lambda}},\label{eq:eff61}
\end{equation}
where $\varphi_{0}$ is another constant.

\section{Equivalence principle violation via quantum field theory}

There exists an extremely elegant way to treat EP violation in a QFT and GR
framework~\cite{don} (for modified gravity, see for example Refs.~\cite{hui} and Ref.\cite{sc} for the generalized uncertainty principle).
The system we want to study consists of an electron with mass $m_{0}$ (the
renormalized mass of the particle when the temperature is zero) in thermal equilibrium with a photon heat bath. The aim of the
analysis is the evaluation of electron's gravitational and inertial mass
in the low-temperature limit (namely, $T\ll m_{0}$). The presence
of a non-zero temperature is fundamental, since calculations clearly
show that $m_{g}=m_{i}$ for $T=0$.

The gravitational and inertial masses  are derived by adopting a Foldy\textendash Wouthuysen
transformation~\cite{fw} on the Dirac equation. Such
a procedure gives the opportunity to study the non-relativistic limit
of particles with spin $\frac{1}{2}$ (i.e. electrons).
In other words, it is possible to derive a Schr\"odinger equation
in which the expression for the mass is easily recognizable.

In order to find a proper expression for $m_i$, one can imagine to switch
an electric field on, so that the Dirac equation which
includes the electromagnetic interaction turns out to be:
\begin{equation}
\left(\cancel{p}-m_{0}-\frac{\alpha}{4\pi^{2}}\cancel{I}\right)\psi=e\Gamma_{\mu}A^{\mu}\psi.\label{eq:92}
\end{equation}
In Eq.~(\ref{eq:92}), $\alpha$ is the fine-structure constant, as usual $\cancel{p}=\gamma^{\mu}p_{\mu}$, with $\gamma^{\mu}$
being the Dirac matrices, $A^{\mu}$ is the electromagnetic four-potential, namely $A^{\mu}=\left(\phi,\mathbf{A}\right)$,
where $\phi$ is the scalar potential and $\mathbf{A}$ the vector potential and the quantity $I_\mu$ is defined as 
\begin{equation}
I_{\mu}=2\int d^{3}k\frac{n_{B}\left(k\right)}{k_{0}}\frac{k_{\mu}}{\omega_{p}k_{0}-\mathbf{p}\cdot\mathbf{k}},\label{eq:89}
\end{equation}
with $k_{\mu}=\left(k_{0},\mathbf{k}\right)$ and where $\omega_{p}$
and $\mathbf{p}$ are connected by:
\begin{equation}
\omega_{p}=\sqrt{m_{0}^{2}+|\mathbf{p}|^{2}}.\label{eq:90}
\end{equation}
In Eq.~(\ref{eq:89}), $n_{B}(k)$ represents the Bose-Einstein
distribution:
\begin{equation}
n_{B}(k)=\frac{1}{e^{\beta k}-1},\label{eq:91}
\end{equation}
where $\beta={1}/{k_{B}T}$, with $k_{B}$ being the Boltzmann constant.
Finally, $\Gamma_{\mu}$ is:
\begin{equation}
\Gamma_{\mu}=\gamma_{\mu}\left(1-\frac{\alpha}{4\pi^{2}}\frac{I_{0}}{E}\right)+\frac{\alpha}{4\pi^{2}}I_{\mu}.\label{eq:94}
\end{equation}
At this point, a Foldy\textendash Wouthuysen transformation converts
Eq.~(\ref{eq:92}) into a Schr\"odinger equation, which reads:
\begin{equation}
i\frac{\partial\psi_{s}}{\partial t}=\left[m_{0}+\frac{\alpha\pi T^{2}}{3m_{0}}+\frac{|\mathbf{p}|^{2}}{2\left(m_{0}+\frac{\alpha\pi T^{2}}{3m_{0}}\right)}+e\phi+\frac{\mathbf{p}\cdot\mathbf{A}+\mathbf{A}\cdot\mathbf{p}}{2\left(m_{0}+\frac{\alpha\pi T^{2}}{3m_{0}}\right)}+\ldots\right]\psi_{s},\label{eq:95}
\end{equation}
from which one extracts the inertial mass:
\begin{equation}
m_{i}=m_{0}+\frac{\alpha\pi T^{2}}{3m_{0}}.\label{eq:96}
\end{equation}
We immediately notice that the difference between the
inertial mass of an electron at finite temperature and $m_{0}$ is
due exclusively to the thermal radiative correction of Eq.~(\ref{eq:96}). 

An analogous reasoning can be performed also for the gravitational mass,
which can be derived in the same way, but starting from a different Dirac equation
that takes into account the gravitational interaction. In a similar circumstance,
one can write:
\begin{equation}
\left(\cancel{p}-m_{0}-\frac{\alpha}{4\pi^{2}}\cancel{I}\right)\psi=\frac{1}{2}h_{\mu\nu}\tau^{\mu\nu}\psi,\label{eq:97}
\end{equation}
where a weak gravitational field is considered (the fluctuation with respect to the background metric are defined by $h_{\mu\nu}$)
and with $\tau^{\mu\nu}$ being the renormalized stress-energy
tensor. In the previous expression, following Refs.~\cite{don},  it is assumed $h_{\mu\nu}~=~2\,\phi_g\,\mathrm{diag}\left(1,1,1,1\right)$, where $\phi_g$ is a gravitational potential.

Once again, a Foldy\textendash Wouthuysen transformation yields another
Schr\"odinger equation:
\begin{equation}
i\frac{\partial\psi_{s}}{\partial t}=\left[m_{0}+\frac{\alpha\pi T^{2}}{3m_{0}}+\frac{|\mathbf{p}|^{2}}{2\left(m_{0}+\frac{\alpha\pi T^{2}}{3m_{0}}\right)}+\left(m_{0}-\frac{\alpha\pi T^{2}}{3m_{0}}\right)\phi_{g}\right]\psi_{s},\label{eq:100}
\end{equation}
from which the identification of the gravitational mass is an easy task:
\begin{equation}
m_{g}=\left(m_{0}-\frac{\alpha\pi T^{2}}{3m_{0}}\right).\label{eq:101}
\end{equation}
Such an outcome implies that there is no difference at
all between $m_{g}$ and $m_{i}$ at zero temperature, because they
both equal the renormalized mass.
This means that only radiative corrections render the violation
of the equivalence principle feasible.

At this point, it is straightforward to check that Eqs.~(\ref{eq:96}) and (\ref{eq:101}) give:
\begin{equation}
\frac{m_{g}}{m_{i}}=1-\frac{2\alpha\pi T^{2}}{3m_{0}^{2}},\label{eq:103}
\end{equation}
in the first-order approximation in $T^{2}$, legitimated by the choice
of evaluating the low-temperature limit of the analyzed quantum system.
The last expression is a direct consequence of the
fact that Lorentz invariance of the finite temperature vacuum is broken, which means that it is possible to define an absolute motion
through the vacuum (i.e. the one at rest with the heat bath). 

Eq.~(\ref{eq:103}) is the core of our argumentation that will be developed below. 
The central result is the violation of EP,
achievable  by means of QFT at finite temperature.
Regarding this point, the question arises
whether $T$ can be inserted into GR  with the aim to reproduce
the same outcome bypassing radiative correction computations.
If this proposal is viable, it should be possible to  develop similar calculations
for several physical frames describing different space-times. 

\section{Equivalence principle violation via modified geodesic equation}

Considering if Eq.~(\ref{eq:103})
can be derived involving exclusively GR properties is the  aim of this section. 
The derivation exhibited below closely follows the original one contained in Ref.~\cite{gasp}.

Let us study the aforementioned procedure to reach Eq.~(\ref{eq:103})
once again, but from a different path. The starting point is the analysis of a charged
test particle of renormalized mass at zero temperature $m_{0}$ in
thermal equilibrium with a photon heat bath in the low-temperature
limit $T\ll m_{0}$. Hence, the dispersion relation is modified by an additional term~\cite{don}:
\begin{equation}
E=\sqrt{m_{0}^{2}+|\mathbf{p}|^{2}+\frac{2}{3}\alpha\pi T^{2}},\label{eq:104}
\end{equation}
which can be easily identified with the first-order correction in
$T^{2}$ that descends from the finite temperature analysis.

At this point, let us introduce the stress-energy tensor $T^{\mu\nu}$
related to the test particle, whose world line can be contained in
a narrow ``world tube'' in which $T^{\mu\nu}$ is non-vanishing. The conservation equation 
for the stress-energy tensor can be integrated over a three-dimensional
hyper-surface $\Sigma$ and defined as:
\begin{equation}
\int_{\Sigma}d^{3}x'\sqrt{-g}T^{\mu\nu}\left(x'\right)=\frac{p^{\mu}p^{\nu}}{E},\label{eq:105}
\end{equation}
where $p^{\mu}$ is the four-momentum and $E=p^{0}$ the energy, given
by:
\begin{equation}
E=\int_{\Sigma}d^{3}x'\sqrt{-g}T^{00}\left(x'\right).\label{eq:106}
\end{equation}
These equations hold in the limit where the world
tube radius goes to zero~\cite{pap}.

A more accurate study~\cite{don} gives the term that should
be viewed as the source of gravity at finite temperature and
in weak-field approximation, which, in the
rest frame of the heat bath, turns out to be:
\begin{equation}
\Xi^{\mu\nu}=T^{\mu\nu}-\frac{2}{3}\alpha\pi\frac{T^{2}}{E^{2}}\delta_{\;\;0}^{\mu}\delta_{\;\;0}^{\nu}T^{00},\label{eq:107}
\end{equation}
where $\Xi^{\mu\nu}$ contains not only the information on the Einstein tensor $G^{\mu\nu}$, but also thermal corrections to it.

Eq. (\ref{eq:107}) is explicitly derived after the choice
of the privileged reference frame at rest with the heat bath, and this fact
produces a Lorentz invariance violation of the finite temperature
vacuum. In fact, in the flat tangent space, one cannot
consider a  Minkowski vacuum anymore, since it is substituted by 
a thermal bath. For this reason, Lorentz group is no longer the symmetry
group of the local tangent space to the Riemannian manifold, even though general covariance still
holds there. The last consideration allows us to proceed with the awareness
that the current situation is slightly different from the usual GR scheme.

However, since the analyzed case deals with weak-field approximation
and quadratic thermal corrections in low-temperature limit, the generalization
of Eq.~(\ref{eq:107}) to a curved space-time can be:
\begin{equation}
\Xi^{\mu\nu}=T^{\mu\nu}-\frac{2}{3}\alpha\pi\frac{T^{2}}{E^{2}}e_{\;\;\hat{0}}^{\mu}e_{\;\;\hat{0}}^{\nu}T^{\hat{0}\hat{0}},\label{eq:108}
\end{equation}
where $e_{\;\;\hat{0}}^{\mu}$ denotes the vierbein field and the hatted indexes are the ones related to the flat tangent space.

Another fundamental assumption has be made to proceed further: effects of temperature on geometry  have not to be taken into account~\cite{gasp}. If the last assertion holds, it is possible to
write the Einstein field equations as:
\begin{equation}
G^{\mu\nu}=\Xi^{\mu\nu},\label{eq:109}
\end{equation}
otherwise other contributions would arise from a relativistic investigation
on temperature (because also $T$ would have an influence on space-time
structure), but for our purposes they can be safely neglected.

If now we employ the Bianchi identity
(namely, $\nabla_{\nu}G^{\mu\nu}=0$), it is straightforward to check that:
\begin{equation}
\nabla_{\nu}T^{\mu\nu}=\nabla_{\nu}\left(\frac{2}{3}\alpha\pi\frac{T^{2}}{E^{2}}e_{\;\;\hat{0}}^{\mu}e_{\;\;\hat{0}}^{\nu}T^{\hat{0}\hat{0}}\right),\label{eq:110}
\end{equation}
which can be rewritten as:
\begin{equation}
\partial_{\nu}\left(\sqrt{-g}T^{\mu\nu}\right)+\Gamma_{\nu\alpha}^{\;\;\;\;\mu}\sqrt{-g}T^{\alpha\nu}=\partial_{\nu}\left(\sqrt{-g}\frac{2}{3}\alpha\pi\frac{T^{2}}{E^{2}}e_{\;\;\hat{0}}^{\mu}e_{\;\;\hat{0}}^{\nu}T^{\hat{0}\hat{0}}\right)+\frac{2}{3}\alpha\pi\Gamma_{\nu\alpha}^{\;\;\;\;\mu}\sqrt{-g}\frac{T^{2}}{E^{2}}e_{\;\;\hat{0}}^{\mu}e_{\;\;\hat{0}}^{\nu}T^{\hat{0}\hat{0}}.\label{eq:111}
\end{equation}
By denoting  $\overset{.}{x}^\mu\equiv dx^\mu/ds$, it can be shown~\cite{gasp} that Eq.~(\ref{eq:111}) is equal to:
\begin{equation}
\overset{..}{x}^{\mu}+\Gamma_{\alpha\nu}^{\;\;\;\;\mu}\overset{.}{x}^{\alpha}\overset{.}{x}^{\nu}=\frac{d}{ds}\left(\frac{2}{3}\alpha\pi\frac{T^{2}}{mE}e_{\;\;\hat{0}}^{\mu}\right)+\frac{2}{3}\alpha\pi\frac{T^{2}}{m^{2}}\Gamma_{\alpha\nu}^{\;\;\;\;\mu}e_{\;\;\hat{0}}^{\alpha}e_{\;\;\hat{0}}^{\nu},\label{eq:114}
\end{equation}
which can be cast into another form by using the fact that:
\begin{equation}
E=m\overset{.}{x}^{\hat{0}}=m\overset{.}{x}^{\rho}e_{\rho}^{\;\;\hat{0}}.\label{eq:115}
\end{equation}
This substitution finally gives:
\begin{equation}
\overset{..}{x}^{\mu}+\Gamma_{\alpha\nu}^{\;\;\;\;\mu}\overset{.}{x}^{\alpha}\overset{.}{x}^{\nu}=\frac{2}{3}\alpha\pi T^{2}\left[\frac{\overset{.}{x}^{\nu}\partial_{\nu}e_{\;\;\hat{0}}^{\mu}}{mE}-\frac{e_{\;\;\hat{0}}^{\mu}\left(\overset{..}{x}^{\nu}e_{\nu}^{\;\;\hat{0}}+\overset{.}{x}^{\nu}\overset{.}{x}^{\beta}\partial_{\beta}e_{\nu}^{\;\;\hat{0}}\right)}{E^{2}}+\frac{\Gamma_{\alpha\nu}^{\;\;\;\;\mu}e_{\;\;\hat{0}}^{\alpha}e_{\;\;\hat{0}}^{\nu}}{m^{2}}\right].\label{eq:116}
\end{equation}
Eq.~(\ref{eq:116}) represents a generalization of the geodesic equation to the 
case in which the temperature is non-vanishing. 

\subsection{Application to the Schwarzschild metric}

We are ready to analyze Eq.~(\ref{eq:116}) in the context of the Schwarzschild solution. To this aim, we can write the metric tensor as:
\begin{equation}
g_{\mu\nu}=\mathrm{diag}\left(e^\nu,-e^\lambda,-r^2,-r^2\mathrm{sin}^2\theta\right), \qquad e^{\nu}=e^{-\lambda}=1-2\phi=1-\frac{2M}{r}.\label{eq:117}
\end{equation}
Moreover, let us recall that $\partial_{t}\phi=0$
and let us assume that only radial motion is considered ($\overset{.}{\vartheta}=\overset{.}{\varphi}=0$).
The vierbeins for the metric of Eq.~(\ref{eq:117}) are:
\begin{equation}
e_{\;\;\hat{0}}^{0}=e^{-\frac{\nu}{2}};\;\;\;e_{\;\;\hat{1}}^{1}=e^{-\frac{\lambda}{2}}.\label{eq:119}
\end{equation}
In addition, we report the expression of the non-vanishing Christoffel symbols:
\begin{equation}
\Gamma_{00}^{\;\;\;\;0}=0;\;\;\;\Gamma_{01}^{\;\;\;\;0}=\frac{\nu'}{2};\;\;\;\Gamma_{11}^{\;\;\;\;0}=0;\;\;\;\Gamma_{00}^{\;\;\;\;1}=\frac{\nu'}{2}\,e^{2\nu};\;\;\;\Gamma_{01}^{\;\;\;\;1}=0;\;\;\;\Gamma_{11}^{\;\;\;\;1}=-\frac{\nu'}{2},\label{eq:120}
\end{equation}
where $\nu=\mathrm{ln}\left(1-2\phi\right)$ and $\nu'={d\nu}/{dr}$.
The geodesic equation for $\mu=0$ is:
\begin{equation}
\overset{..}{t}+\nu'\overset{.}{r}\overset{.}{t}=-\frac{2}{3}\alpha\pi T^{2}\left[\frac{\overset{.}{r}\nu'}{2mE}+\frac{\overset{..}{t}+\frac{\overset{.}{r}\overset{.}{t}\nu'}{2}}{E^{2}}e^{\frac{\nu}{2}}\right]e^{-\frac{\nu}{2}},\label{eq:122}
\end{equation}
but if one recalls that $E=m\overset{.}{x}^{\hat{0}}=m\overset{.}{x}^{\alpha}e_{\alpha}^{\;\;\hat{0}}=m\,\overset{.}{t}\,e^{{\nu}/{2}}$,
Eq.~(\ref{eq:122}) can be once again manipulated to obtain:
\begin{equation}
\overset{..}{t}+\nu'\overset{.}{r}\overset{.}{t}=-\frac{2\alpha\pi T^{2}}{3E^{2}}\left(\overset{..}{t}+\nu'\overset{.}{r}\overset{.}{t}\right),\label{eq:123}
\end{equation}
and since
$\overset{.}{\nu}=\nu'\overset{.}{r}$,
the final relation for the temporal part will be:
\begin{equation}
\left(1+\frac{2\alpha\pi T^{2}}{3E^{2}}\right)\left(\overset{..}{t}+\overset{.}{\nu}\overset{.}{t}\right)=0.\label{eq:125}
\end{equation}
The radial contribution can be computed involving Eq.~(\ref{eq:116}) for $\mu=1$:
\begin{equation}
\overset{..}{r}+\frac{\nu'}{2}\left(\overset{.}{t}^{2}e^{2\nu}-\overset{.}{r}^{2}\right)=\frac{2\alpha\pi T^{2}}{3m^{2}}\frac{e^{\nu}\nu'}{2},\label{eq:126}
\end{equation}
which can be reformulated in a different fashion:
\begin{equation}
\overset{..}{r}+\frac{\nu'}{2}\left(\overset{.}{t}^{2}e^{\nu-\lambda}-\overset{.}{r}^{2}-\frac{2\alpha\pi T^{2}}{3m^{2}}e^{-\lambda}\right)=0.\label{eq:127}
\end{equation}
Eqs. (\ref{eq:125}) and (\ref{eq:127}) constitute a coupled system of
differential equations, which, in general, can be quite difficult to
solve. In this case, however,
simple calculations lead to a handy relation between $\overset{.}{t}^{2}$
and $\overset{.}{r}^{2}$ which can be adopted to find the
desired outcome.

In fact, Eq.~(\ref{eq:127}) can be cast in the form:
\begin{equation}
2\overset{..}{r}-\overset{.}{r}^{2}\nu'+\overset{.}{t}^{2}\nu'e^{2\nu}-\frac{2\alpha\pi T^{2}}{3m^{2}}\nu'e^{\nu}=0.\label{eq:128}
\end{equation}
The previous expression can be written as:
\begin{equation}
e^{\nu}\frac{d}{dr}\left(e^{\lambda}\overset{.}{r}^{2}-e^{\nu}\overset{.}{t}^{2}-\frac{2\alpha\pi T^{2}}{3m^{2}}\nu\right)=0,\label{eq:135}
\end{equation}
which implies:
\begin{equation}
e^{\lambda}\overset{.}{r}^{2}-e^{\nu}\overset{.}{t}^{2}-\frac{2\alpha\pi T^{2}}{3m^{2}}\nu=\mathrm{const}.\label{eq:136}
\end{equation}
The constant can be determined from the condition of normalization
on four-velocity in the limit $\phi\rightarrow0$. Such a requirement
is possible due to the hypothesis made above, namely the independence
of the geometric structure of temperature. Hence,
normalization of $\overset{.}{x}^{\mu}$ implies:
\begin{equation}
\overset{.}{x}^{\mu}\overset{.}{x}_{\mu}=g_{\mu\nu}\overset{.}{x}^{\mu}\overset{.}{x}^{\nu}=1,\label{eq:137}
\end{equation}
or explicitly:
\begin{equation}
e^{\lambda}\overset{.}{r}^{2}-e^{\nu}\overset{.}{t}^{2}=-1,\label{eq:138}
\end{equation}
because angles are fixed quantities.

In the limit of vanishing gravitational field (namely, $\nu,\lambda\rightarrow 0$ as $r\rightarrow\infty$),
Eq.~(\ref{eq:138})  reduces to:
\begin{equation}
\overset{.}{r}_{\infty}^{2}-\overset{.}{t}_{\infty}^{2}=-1.\label{eq:139}
\end{equation}
Such an expression clearly fits also Eq.~(\ref{eq:136}), and thus we have:
\begin{equation}
e^{\lambda}\overset{.}{r}^{2}-e^{\nu}\overset{.}{t}^{2}-\frac{2\alpha\pi T^{2}}{3m^{2}}\nu=-1.\label{eq:140}
\end{equation}
At this point, let us invoke the weak-field approximation. Within this regime and by virtue of Eq.~(\ref{eq:140}), it is immediate to find that Eq.~(\ref{eq:127}) results modified as:
\begin{equation}
\overset{..}{r}=-\frac{M}{r^{2}}\left(1-\frac{2\alpha\pi T^{2}}{3m^{2}}\right),\label{eq:145}
\end{equation}
and if one considers the first-order approximation in $T^{2}$ just like in
QFT considerations, the outcome is:
\begin{equation}\nonumber
\frac{m_{g}}{m_{i}}=1-\frac{2\alpha\pi T^{2}}{3m_{0}^{2}},\label{eq:146}
\end{equation}
which is exactly Eq.~(\ref{eq:103}).

With the last result, the equivalence between the QFT approach and the one formulated in Ref.~\cite{gasp} has been clarified, even though they differ
in some aspects. However, both of the two investigations base their
development on a finite-temperature analysis, that enables the emergence
of the radiative correction in the ratio $ª{m_{g}}/{m_{i}}$.

As anticipated in the previous section, the modified geodesic equation
is a general result, and, as such, it can be applied to any physical system;
the only quantity, required for the  calculations, is the metric tensor.

\subsection{Application to a generic diagonal metric}

In what follows, we derive a general prescription to extract a key relation in the most general
situation of a diagonal metric tensor written in spherical coordinates, whose terms depend exclusively on the modulus of $r$. In other words, we can apply this method whenever $g_{\mu\nu}$ can be cast as:
\begin{equation}
g_{\mu\nu}=\mathrm{diag}\left(A(r),-\frac{1}{B(r)},-r^2,-r^2\mathrm{sin}^2\theta\right).
\end{equation}
As usual, Christoffel symbols and vierbein fields must be expressed
in order to reach the desired outcome:
\begin{equation}
e_{\;\;\hat{0}}^{0}=\left(A\right)^{-\frac{1}{2}};\;\;\;e_{\;\;\hat{1}}^{1}=\left(B\right)^{\frac{1}{2}},\label{eq:263}
\end{equation}
\begin{equation}
\Gamma_{00}^{\;\;\;\;0}=0;\;\;\;\Gamma_{01}^{\;\;\;\;0}=\frac{\partial_{r}A(r)}{2A(r)};\;\;\;\Gamma_{11}^{\;\;\;\;0}=0;\;\;\;\Gamma_{00}^{\;\;\;\;1}=\frac{B(r)}{2}\partial_{r}A(r);\;\;\;\Gamma_{01}^{\;\;\;\;1}=0;\;\;\;\Gamma_{11}^{\;\;\;\;1}=\frac{B(r)}{2}\partial_{r}\left(\frac{1}{B(r)}\right).\label{eq:a22}
\end{equation}
As before, we shall assume $\overset{.}{\theta}=\overset{.}{\varphi}=0$.

The first expression to be analyzed is related to the temporal coordinate:
\begin{equation}
\overset{..}{t}+\overset{.}{r}\overset{.}{t}\frac{\partial_{r}A}{A}=\frac{2}{3}\alpha\pi T^{2}\left[-\frac{\overset{.}{r}}{2mE}\frac{\partial_{r}A}{\left(A\right)^{\frac{3}{2}}}-\frac{1}{E^{2}}\left(\overset{..}{t}+\overset{.}{r}\overset{.}{t}\frac{\partial_{r}A}{2A}\right)\right],\label{eq:264}
\end{equation}
but since $E=m\,\overset{.}{t}\,e_{0}^{\;\;\hat{0}}=m\,\overset{.}{t}\,\sqrt{A}$,
Eq.~(\ref{eq:264}) can be rewritten as follows: 
\begin{equation}
\overset{..}{t}+\overset{.}{r}\overset{.}{t}\frac{\partial_{r}A}{A}=-\frac{2\alpha\pi T^{2}}{3E^{2}}\left[\overset{..}{t}+\overset{.}{r}\overset{.}{t}\frac{\partial_{r}A}{A}\right],\label{eq:265}
\end{equation}
or equivalently:
\begin{equation}
\left(1+\frac{2\alpha\pi T^{2}}{3E^{2}}\right)\left(\overset{..}{t}+\overset{.}{r}\overset{.}{t}\frac{\partial_{r}A}{A}\right)=0,\label{eq:266}
\end{equation}
which returns Eq.~(\ref{eq:125}) with the choice $A=e^{\nu}$.

The radial equation is instead given by:
\begin{equation}
\overset{..}{r}+\overset{.}{r}^{2}\frac{B}{2}\partial_{r}\left(\frac{1}{B}\right)+\overset{.}{t}^{2}\frac{B\partial_{r}A}{2}=\frac{2\alpha\pi T^{2}}{3m^{2}}\frac{B}{2}\frac{\partial_{r}A}{A},\label{eq:267}
\end{equation}
or:
\begin{equation}
\overset{..}{r}+\frac{B}{2}\left[\overset{.}{t}^{2}\partial_{r}A+\overset{.}{r}^{2}\partial_{r}\left(\frac{1}{B}\right)-\frac{2\alpha\pi T^{2}}{3m^{2}}\frac{\partial_{r}A}{A}\right]=0,\label{eq:268}
\end{equation}
which exactly yields Eq.~(\ref{eq:127}) with the proper values of
$A$ and $B$.

Let us consider
Eqs.~(\ref{eq:266}) and (\ref{eq:268}) to determine the unknown
quantities. From the first relation, it is easy to see that:
\begin{equation}
\frac{d}{ds}\left(\mathrm{ln}\,\overset{.}{t}\right)=-\frac{d}{ds}\left(\mathrm{ln}\,A\right),\label{eq:269}
\end{equation}
and hence:
\begin{equation}
\overset{.}{t}=\frac{1}{A}.\label{eq:270}
\end{equation}
For what concerns the radial equation, it can be noticed that all contributions can be written in terms of a total derivative, as in the previous case of the Schwarzschild geometry:
\begin{equation}
\frac{B}{2}\frac{d}{dr}\left[\frac{\overset{.}{r}^{2}}{B}-\overset{.}{t}^{2}A-\frac{2\alpha\pi T^{2}}{3m^{2}}\mathrm{ln}\left(A\right)\right]=0.\label{eq:276}
\end{equation}
The solution of the above equation (assuming the same normalization condition as before) is:
\begin{equation}
\frac{\overset{.}{r}^{2}}{B}-\overset{.}{t}^{2}A-\frac{2\alpha\pi T^{2}}{3m^{2}}\mathrm{ln}\left(A\right)=-1,\label{eq:277}
\end{equation}
which is the generalization of Eq.~(\ref{eq:140}).

The above expression can be reformulated as:
\begin{equation}
\overset{.}{r}^{2}=B\left(\frac{1}{A}+\frac{2\alpha\pi T^{2}}{3m^{2}}\mathrm{ln}\left(A\right)-1\right).\label{eq:278}
\end{equation}
Finally, the insertion of Eqs.~(\ref{eq:270}) and (\ref{eq:278})
into Eq.~(\ref{eq:268}) gives:
\begin{equation}
\overset{..}{r}=-\frac{B}{2}\left[\frac{\partial_{r}A}{A^{2}}-\left(\frac{1}{A}-1\right)\frac{\partial_{r}B}{B}-\frac{2\alpha\pi T^{2}}{3m^{2}}\left(\frac{\partial_{r}A}{A}+\frac{\mathrm{ln}\left(A\right)\partial_{r}B}{B}\right)\right].\label{eq:279}
\end{equation}
Eq.~(\ref{eq:279}) represents the final point for the current analysis, which fits any metric tensor that is diagonal, written in spherical coordinate system and whose components  have only a dependence on $r$. 

\subsection{Application to the Brans-Dicke case}

Before applying of the above formalism to the spherical metric derived from the Brans-Dicke theory, let
us remind some basic properties. 
The physical quantities  that we need  are the Christoffel symbols and the tetrads, which
once again are easy to calculate being $g_{\mu\nu}$  diagonal (see Eq.~(\ref{eq:eff57})):
\begin{equation}
e_{\;\;\hat{0}}^{0}=e^{-\frac{v}{2}};\;\;\;e_{\;\;\hat{1}}^{1}=e^{-\frac{u}{2}},\label{eq:eff263}
\end{equation}
\begin{equation}
\Gamma_{00}^{\;\;\;\;0}=0;\;\;\;\Gamma_{01}^{\;\;\;\;0}=\frac{v'}{2};\;\;\;\Gamma_{11}^{\;\;\;\;0}=0;\;\;\;\Gamma_{00}^{\;\;\;\;1}=\frac{v'}{2}\,e^{v-u};\;\;\;\Gamma_{01}^{\;\;\;\;1}=0;\;\;\;\Gamma_{11}^{\;\;\;\;1}=-\frac{u'}{2}.
\end{equation}
Explicit formulas for $u$ and $v$ are:
\begin{equation}
v=2\alpha_{0}+\frac{2}{\lambda}\mathrm{ln}\left(\frac{1-\frac{B}{r}}{1+\frac{B}{r}}\right), \qquad u=2\beta_{0}+4\,\mathrm{ln}\left(1+\frac{B}{r}\right)+\frac{2}{\lambda}\left(\lambda-C-1\right)\mathrm{ln}\left(\frac{1-\frac{B}{r}}{1+\frac{B}{r}}\right),\label{eq:eff264}
\end{equation}
where it is clear that the information related to $\omega$ is hidden in the constants, whereas the mass of
the gravitational source is contained in the parameter $B$.

After this  digression, it is possible to evaluate the temporal
differential equation:
\begin{equation}
\overset{..}{t}+\overset{.}{r}\overset{.}{t}v'=\frac{2}{3}\alpha\pi T^{2}\left[-\frac{\overset{.}{r}v'e^{-\frac{v}{2}}}{2mE}+\frac{\overset{..}{t}+\overset{.}{r}\overset{.}{t}\frac{v'}{2}}{E^{2}}\right]\,.\label{eq:eff266}
\end{equation}
Being $E=m\,\overset{.}{t}\,e^{{v}/{2}}$, the previous relation can be reformulated
as:
\begin{equation}
\left[1+\frac{2\alpha\pi T^{2}}{3E^{2}}\right]\left(\overset{..}{t}+\overset{.}{v}\overset{.}{t}\right)=0,\label{eq:eff267}
\end{equation}
which is formally equal to the expression obtained for Schwarzschild,
but in this case $v$ has a  different meaning.

The radial equation is similar to  Eq.~(\ref{eq:127}):
\begin{equation}
\overset{..}{r}+\overset{.}{r}^{2}\frac{u'}{2}+\overset{.}{t}^{2}\frac{v'e^{v-u}}{2}=\frac{2\alpha\pi T^{2}}{3m^{2}}\frac{v'e^{-u}}{2},\label{eq:eff268}
\end{equation}
or equivalently:
\begin{equation}
\overset{..}{r}+\frac{v'}{2}\left[\overset{.}{t}^{2}e^{v-u}+\overset{.}{r}^{2}\frac{u'}{v'}-\frac{2\alpha\pi T^{2}}{3m^{2}}e^{-u}\right]=0.\label{eq:eff269}
\end{equation}
If $u'=-v'$ one exactly obtains the above results of the  Schwarzschild solution.

However, this is not the final expression for the Brans-Dicke case. In fact, Eq.~(\ref{eq:eff269}) can be further simplified 
 adopting  the same method that leads to Eq.~(\ref{eq:140}) in
the previous section. Here, the situation is  similar, and thus:
\begin{equation}
\overset{.}{r}^{2}e^{u}-\overset{.}{t}^{2}e^{v}-\frac{2\alpha\pi T^{2}}{3m^{2}}v=-1.\label{eq:eff270}
\end{equation}
Neglecting higher-order terms with respect to $\varphi$ would exclude interesting
contributions to the ratio ${m_{g}}/{m_{i}}$.
Hence, another way to simplify Eq.~(\ref{eq:eff269}) must be found.
In order to do that, $\overset{.}{t}$ and $\overset{.}{r}$ are the quantities that should be explicitly expressed, since their evolution has not been determined yet. However, this turns
out to be easy by virtue of Eqs.~(\ref{eq:eff267}) and (\ref{eq:eff270}). Indeed, the first one tells that:
\begin{equation}
\frac{\overset{..}{t}}{\overset{.}{t}}=-\frac{d}{ds}v,\label{eq:eff271}
\end{equation}
but it is evident that ${\overset{..}{t}}/{\overset{.}{t}}={d}/{ds}\left[\mathrm{ln}\,\overset{.}{t}\right]$,
and for this reason the differential equation can be immediately solved:
\begin{equation}
\overset{.}{t}=e^{-v}.\label{eq:eff272}
\end{equation}
Thanks to Eq.~(\ref{eq:eff272}), it is possible to find an expression
also for $\overset{.}{r}^{2}$:
\begin{equation}
\overset{.}{r}^{2}=\left(e^{-v}+\frac{2\alpha\pi T^{2}}{3m^{2}}v-1\right)e^{-u},\label{eq:eff273}
\end{equation}
and with these two expressions, Eq.~(\ref{eq:eff269}) can be rewritten
as:
\begin{equation}
\overset{..}{r}=-\frac{v'}{2}\left[e^{-v}\left(1+\frac{u'}{v'}\right)-\frac{u'}{v'}-\frac{2\alpha\pi T^{2}}{3m^{2}}\left(1-\frac{u'}{v'}v\right)\right]e^{-u}.\label{eq:eff274}
\end{equation}
Nevertheless, the above-mentioned quantities are not so easy to handle
within this framework, since their expression is rather convoluted.
However, a further step can link $u'$ with $v'$ starting from
Eq.~(\ref{eq:eff264}). One then has:
\begin{equation}
v'=-\frac{4B}{\lambda}\left(\frac{1}{B^{2}-r^{2}}\right), \qquad u'=-\frac{4B}{\lambda}\left(\frac{\frac{\lambda B}{r}-C-1}{B^{2}-r^{2}}\right).\label{eq:eff276}
\end{equation}
With these definitions, the   Brans-Dicke constants appear
in Eq.~(\ref{eq:eff274}). Such a statement 
is non-trivial, because the shift between gravitational and inertial
mass will depend on $\omega$.

Apart from the previous prediction, it can be easily observed that:
\begin{equation}
\frac{u'}{v'}=\frac{\lambda B}{r}-C-1,\label{eq:eff278}
\end{equation}
and as a consequence:
\begin{equation}
\overset{..}{r}=-\frac{v'}{2}\left\{ 1+\left(e^{-v}-1\right)\left(\frac{\lambda B}{r}-C\right)-\frac{2\alpha\pi T^{2}}{3m^{2}}\left[1+v-\left(\frac{\lambda B}{r}-C\right)v\right]\right\} e^{-u}.\label{eq:eff279}
\end{equation}
Two crucial aspects must be highlighted before proceding. We want to stress that Eq.~(\ref{eq:eff279}) can be directly
obtained by the most general one, Eq.~(\ref{eq:279}), with the substitution $A\left(r\right)=e^{v}$
and $B\left(r\right)=e^{-u}$, as expected. The other remark is related to the fact that
scalar-tensor theories do not possess
only one fundamental field.
For this reason, in principle, there should be another contribution
in Eq.~(\ref{eq:eff279}) that takes into account the presence of the
stress-energy tensor related to the scalar field $\varphi$. However,
the influence of $T$ on $\varphi$ has not
been considered here.


Going back to Eq.~(\ref{eq:eff279}), one can observe that there is
not only the radiative correction to the ratio ${m_{g}}/{m_{i}}$,
but also another contribution which exclusively depends on $\omega$ and that
correctly vanishes in the limit $\omega\rightarrow\infty$, that is when GR is recovered.
Moreover, the evaluation of the second quantity of Eq.~(\ref{eq:eff279})
represents an important opportunity to put a lower bound to the parameter
of the Brans-Dicke theory. In fact, if one imposes that $|({m_{g}-m_{i}})/{m_{i}}|<10^{-14}$~\cite{bae} and studies the factor
$\left(e^{-v}-1\right)\left({\lambda B}/{r}-C\right)$ in the
first-order approximation in $\phi_{g}$ casting radiative corrections aside momentarily, it is possible to easily constrain $\omega$. 
In order to perform that, we need the quantities of the Brans-Dicke theory to be written in terms of the free parameter of the model and in the weak-field limit~\cite{fuj,bar}:
\begin{equation}
\alpha_{0}=\beta_{0}=0;\;\;C=-\frac{1}{2+\omega};\;\;B=\frac{GM\lambda}{2};\;\;\lambda=\sqrt{\frac{2\omega+3}{2\omega+4}}.\label{eq:299}
\end{equation}
It is easy to put $B$ in terms of $\phi_{g}$:
\begin{equation}
B=\frac{\lambda r\phi_{g}}{2},\label{eq:300}
\end{equation}
and to expand the function $e^{-v}$:
\begin{equation}
e^{-v}=\left(\frac{1-\frac{\lambda\phi_{g}}{2}}{1+\frac{\lambda\phi_{g}}{2}}\right)^{-\frac{2}{\lambda}}\sim1+2\phi_{g}.\label{eq:301}
\end{equation}
Since we neglect higher-order terms 
of $\phi_{g}$, the examined factor is simply:
\begin{equation}
\frac{2\phi_{g}}{2+\omega},\label{eq:302}
\end{equation}
and thus from ${2\phi_{g}}/({2+\omega})<10^{-14}$, we get:
\begin{equation}
\omega>\frac{2GM}{r}\cdot10^{14},\label{eq:305}
\end{equation}
which is the final expression for the lower bound of the 
 Brans-Dicke parameter  in the weak-field approximation.

A similar result is easy to achieve only if weak-field approximation
is performed, otherwise the complete dependence of constants $\lambda$
and $C$ with respect to $\omega$ would have been more difficult
to handle. For instance, let us consider the gravitational field of the Earth by recalling that
$M_{\oplus}=5.97\cdot10^{24}\;Kg;\;\;\;R_{\oplus}=6.37\cdot10^{6}\;m.$
It is immediate to achieve:
\begin{equation}
\omega>1.40\cdot10^{5},\label{eq:308}
\end{equation}
that is similar to a bound recently  experimentally obtained~\cite{wy}, which gives $\omega>3\cdot10^{5}$.
For the sake of completeness, it is useful to look at a table that
contains a prediction of the most reliable bounds for $\omega$~\cite{ap}.
\vspace{-5mm}
\begin{table}[H]
\centering
\begin{tabular}{|c|c|c|c|}
\hline 
Detector & System & Specification & Expected bound on $\omega$\tabularnewline
\hline 
\hline 
aLIGO & $\left(1.4+5\right)M_{\odot}$ & 100 Mpc & $\sim100$\tabularnewline
\hline 
ET & $\left(1.4+5\right)M_{\odot}$ & 100 Mpc & $\sim10^{5}$\tabularnewline
\hline 
ET & $\left(1.4+2\right)M_{\odot}$ & 100 Mpc & $\sim10^{4}$\tabularnewline
\hline 
eLISA & $\left(1.4+400\right)M_{\odot}$ & SNR=10 & $\sim10^{4}$\tabularnewline
\hline 
LISA & $\left(1.4+400\right)M_{\odot}$ & SNR=10 & $\sim10^{5}$\tabularnewline
\hline 
DECIGO & $\left(1.4+10\right)M_{\odot}$ & SNR=10 & $\sim10^{6}$\tabularnewline
\hline 
Cassini & Solar System &  & $\sim10^{4}$\tabularnewline
\hline 
\end{tabular}

\caption{This table includes expected outcomes of experimental observations,
in addition to a known bound deduced by the probe Cassini through
the analysis of the Solar System.}

\end{table}

\section{Discussion and Conclusions}
In this paper, we have analyzed the EP violation  triggered by the presence of a non-vanishing temperature. We have briefly summarized the procedure that firstly led to such an outcome by adopting QFT techniques. After that, we have shown how the same result can be achieved without relying on loop computations, but rather focusing the attention on the modification of the geodesic equation for $T\neq 0$. With this simplified treatment, it is possible to study not only the space-time described by the Schwarzschild solution in GR, but also other physical environments, even the ones arising from Extended Theories of Gravity. In this perspective, the Brans-Dicke theory has been examined, and we have highlighted the possibility to put a constraint on the free parameter of this model by resorting to the current data related to the EP.

In particular, we want to recall the result of Eq.~(\ref{eq:308}),
which directly depends on the ratio of the gravitational and
inertial mass, as it could be seen in Eq.~(\ref{eq:305}), where the
factor $10^{14}$ is an immediate consequence of the statement $|({m_{g}-m_{i}})/{m_{i}}|<10^{-14}$.
If experiments were able to reach an even higher
precision, i.e. $10^{-17}$~\cite{na}, one would have $\omega>{2\,\phi_g}\cdot10^{17}$,
instead of Eq.~(\ref{eq:305}). As a consequence, the lower
bound on $\omega$ for the Earth would be $\omega>1.40\cdot10^{8}$,
which far exceeds the expected outcomes exposed in Table 1.

The above approach is applicable to any framework embedded in a curved background\footnote{Note that the analysis of neutrino physics in curved backgrounds is intimately related to the EP violation, as it can be deduced from the works of Refs.~\cite{lc} and references therein.}, and (as already remarked) can be adopted for any Extended Theory of Gravity. As in the case of Brans-Dicke, in principle it can be possible to use the current data on EP to put considerable bounds on the free parameters of the aforementioned gravitational models. In this sense, such a formalism is similar to recent works in which the Casimir-like systems are employed to extract useful  information on the free quantities of Extended Theories of Gravity~\cite{lamb}.

Finally, it must be stressed that the validity of the EP violation, obtained via the modified geodesic equation, is restricted by the requirement of dealing with a weak gravitational field. Indeed, all the derivation based on Ref.~\cite{gasp} is centered on  this limit, beyond which the approach fails to be valid. In order to look at physical scenarios in the presence of strong gravity (in this direction, there is already an interesting work involving the Unruh and Hawking effects~\cite{sw}), it is necessary either to use a full QFT treatment or to generalize the arguments presented above.

\section*{Acknowledgements}
The Authors acknowledges INFN Sez. di Napoli (Iniziative Specifiche QGSKY and TEONGRAV) and  the COST Action CA15117 (CANTATA).

\end{document}